\documentclass[fleqn,usenatbib]{mnras}

\usepackage[T1]{fontenc}
\usepackage[usenames,dvipsnames,table]{xcolor}
\usepackage{ulem}
\DeclareRobustCommand{\VAN}[3]{#2}
\let\VANthebibliography\thebibliography
\def\thebibliography{\DeclareRobustCommand{\VAN}[3]{##3}\VANthebibliography}

\usepackage{graphicx}	
\usepackage{amsmath}	
\usepackage{amssymb}	
\usepackage{appendix}
\usepackage{newtxtext,newtxmath}

\title[Does the secondary white dwarf survive?]{On the fate of the secondary white dwarf in double-degenerate double-detonation Type~Ia supernovae}

\author[R. Pakmor et al.]{R.~Pakmor$^{1}$\thanks{E-mail: rpakmor@mpa-garching.mpg.de}, F.~P.~Callan$^{2}$, C.~E.~Collins$^{3}$, S.~E.~de~Mink$^{1,4,5}$, A.~Holas$^{6,7}$,  \newauthor W.~E.~Kerzendorf$^{8,9}$, M.~Kromer$^{7}$, P.~G.~Neunteufel$^{1}$, John~T.~O'Brien$^{8}$, F.~K.~R\"{o}pke$^{7,6}$, \newauthor A.~J.~Ruiter$^{10}$, I.~R.~Seitenzahl$^{10}$, Luke~J.~Shingles$^3$, S. A. Sim$^{2}$,  S.~Taubenberger$^{1}$\\
$^1$Max-Planck-Institut f\"{u}r Astrophysik, Karl-Schwarzschild-Str. 1, D-85748, Garching, Germany\\
$^2$School of Mathematics and Physics, Queen’s University Belfast, Belfast BT7 1NN, UK\\
$^3$GSI Helmholtzzentrum f\"{u}r Schwerionenforschung, Planckstraße 1, 64291 Darmstadt, Germany\\
$^4$Anton Pannekoek Institute of Astronomy and GRAPPA, Science Park 904, University of Amsterdam, 1098XH Amsterdam, The Netherlands\\
$^5$Center for Astrophysics $|$ Harvard $\&$ Smithsonian, 60 Garden St., Cambridge, MA 02138, USA\\
$^6$Zentrum f\"{u}r Astronomie der Universit\"{a}t Heidelberg, Institut für Theoretische Astrophysik, Philosophenweg 12, D-69120 Heidelberg, Germany\\
$^7$Heidelberg Institute for Theoretical Studies, Schloss-Wolfsbrunnenweg 35,    69118 Heidelberg, Germany\\
$^8$Department of Physics and Astronomy, Michigan State University, East Lansing, MI 48824, USA\\
$^9$Department of Computational Mathematics, Science, and Engineering, Michigan State University, East Lansing, MI 48824, USA\\
$^{10}$School of Science, University of New South Wales, Canberra, ACT 2600, Australia\\
}

\date{Accepted XXX. Received YYY; in original form ZZZ}

\pubyear{2022}

\begin{document}
\label{firstpage}
\pagerange{\pageref{firstpage}--\pageref{lastpage}}
\maketitle

\begin{abstract}
The progenitor systems and explosion mechanism of Type Ia supernovae are still unknown. Currently favoured progenitors include double-degenerate systems consisting of two carbon-oxygen white dwarfs with thin helium shells. In the double-detonation scenario, violent accretion leads to a helium detonation on the more massive primary white dwarf that turns into a carbon detonation in its core and explodes it. We investigate the fate of the secondary white dwarf, focusing on changes of the ejecta and observables of the explosion if the secondary explodes as well rather than survives. We simulate a binary system of a $1.05\,\mathrm{M_\odot}$ and a $0.7\,\mathrm{M_\odot}$ carbon-oxygen white dwarf with $0.03\,\mathrm{M_\odot}$ helium shells each. We follow the system self-consistently from inspiral to ignition, through the explosion, to synthetic observables. We confirm that the primary white dwarf explodes self-consistently. The helium detonation around the secondary white dwarf, however, fails to ignite a carbon detonation. We restart the simulation igniting the carbon detonation in the secondary white dwarf by hand and compare the ejecta and observables of both explosions. We find that the outer ejecta at $v~>~15\,000$\,km\,s$^{-1}$ are indistinguishable. Light curves and spectra are very similar until $\sim~40\,\mathrm{d}$ after explosion and the ejecta are much more spherical than violent merger models. The inner ejecta differ significantly slowing down the decline rate of the bolometric light curve after maximum of the model with a secondary explosion by $\sim20$ per cent. We expect future synthetic 3D nebular spectra to confirm or rule out either model.
\end{abstract}




\section{Introduction}

The progenitor systems and the explosion mechanism of normal Type Ia supernovae are still unknown \citep{Maoz2014, Livio2018, Ruiter2020}. There is general agreement only that they are thermonuclear explosions of carbon-oxygen white dwarfs with masses $\gtrsim 0.8\,\mathrm{M_\odot}$ in close binary systems. Somehow the interaction with the companion star directly (via accretion induced dynamical effects) or indirectly (by growing the  white dwarf to the Chandrasekhar mass) likely triggers the thermonuclear explosion. The nature of the companion star, a white dwarf or an ordinary non-degenerate star, and the physical mechanism that causes the explosion remain open questions.

Binary systems of two white dwarfs (so-called double-degenerate systems) are good candidates for the progenitor systems of normal Type Ia supernovae \citep{Branch1993}. Their rates are consistent with the observed rate of normal Type Ia supernovae \citep{Ruiter2009} within the uncertainties. A binary system of two white dwarfs is also easily able to explain the lack of any pre-explosion detection of the progenitor system \citep{Weidong2011,Kelly2014}, the lack of hydrogen in any early or late spectra of normal Type Ia supernovae \citep{Lundqvist2015,Maguire2016,Tucker2020}, the clear lack of signatures from interaction of the explosion with circumstellar material for many normal Type Ia supernovae \citep{Margutti2014,Ferretti2017}, as well as the lack of any surviving companion brighter than about solar luminosity \citep[see, e.g.][]{Schaefer2012, GonzalezHernandez2012}. \citet{Kerzendorf2018} has ruled out a blue survivor for the nearby remnant of SN~1006 with little extinction down to $\approx0.01L_\odot$.

Idealised models in which an isolated sub-Chandrasekhar-mass carbon-oxygen white dwarf is artificially ignited at its centre generally agree well with observations with only few systematic differences \citep{Sim2010, Woosley2011, Blondin2017, Shen2018, Shen2021RT}. They are arguably a reasonable simplification of double-degenerate scenarios in which only the primary (more massive) white dwarf explodes \citep{Pakmor2013}. One systematic difference is that these sub-Chandrasekhar-mass models decline faster after maximum brightness than observed light curves. While the faster decline in the $B$ band depends on the radiative-transfer treatment \citep{Sim2010,Shen2021}, the faster decline of the bolometric light curves seems to be a more fundamental problem \citep{Kushnir2020}. This possibly means that the ejecta mass of the toy models is too low to explain more slowly declining events \citep{Stritzinger2006,Scalzo2014,Scalzo2019}.

Double-degenerate systems are also very attractive progenitor candidates because they seem to reproduce the brightness distribution of normal Type Ia supernovae under the assumption that the brightness of the explosion is set only by the mass of the primary white dwarf \citep{Ruiter2013, Sato2016}. They also reproduce some of the correlations between the observed silicon line velocity and the brightness of the explosion \citep{Shen2021RT}. Similarly they potentially explain some of the observed correlations between the brightness of observed Type Ia supernovae and properties of their host galaxies via the age of the progenitor system \citep{Kelly2010, Childress2013}.

However, modelling double-degenerate systems becomes much more complicated when we go beyond idealised models and include the ignition. In the violent mergers scenario \citep{Pakmor2010} when the secondary white dwarf is about to be destroyed the interaction of the debris of the secondary white dwarf with the primary white dwarf directly ignites a carbon detonation on the surface of the primary white dwarf. The following explosion is more asymmetric than normal Type Ia supernovae \citep{Pakmor2012b, Bulla2016}.

Currently it seems more plausible that the explosion ignites via the double-detonation mechanism \citep{Livne1990,Fink2010}. In its modern version, unstable dynamical accretion of helium from the secondary white dwarf just prior to the merger of the binary heats up the helium shell on the primary white dwarf. Eventually, dynamical instabilities from the interaction between the accretion stream and the helium shell of the primary white dwarf lead to a thermonuclear runaway and a helium detonation ignites on the surface of the primary white dwarf. This helium detonation burns the shell around the primary white dwarf and sends a shockwave into its core. The shockwave then converges in a single point in the carbon-oxygen core of the primary white dwarf where it ignites a carbon detonation that explodes the whole primary white dwarf \citep{Guillochon2010,Pakmor2013,Boos2021,Shen2021}. However, there are also recent simulations in which the helium detonation does not ignite a carbon detonation in the primary white dwarf \citep{Roy2022}.

In contrast to the violent merger scenario \citep{Pakmor2012b}, in the double-detonation scenario the secondary white dwarf is generally assumed to survive, because it is still completely intact when the primary white dwarf explodes \citep{Pakmor2013}. This is potentially in conflict with the limited number of candidates for surviving secondary white dwarfs in our neighbourhood \citep{Shen2018b} and the lack of any known fast-moving white dwarfs in nearby Type Ia supernova remnants \citep{Shields2022}.

Here we revisit this scenario, i.e. the helium-ignited explosion of a double-degenerate system with a focus on the fate of the secondary white dwarf. In particular we investigate the possibility that the secondary white dwarf may explode via the same double-detonation mechanism as the primary white dwarf \citep{Pakmor2021}. A similar explosion of the secondary white dwarf has previously been found for massive helium white dwarf companions \citep{Papish2015}. We employ three-dimensional (3D) hydrodynamical simulations with a fully coupled nuclear reaction network to simulate the binary system. We compare a new scenario in which both white dwarfs explode via the double-detonation scenario with the scenario where only the primary white dwarf explodes and the secondary white dwarf survives. We analyse and interpret the differences in the ejecta of the explosions and their synthetic observables.

We describe the simulation codes we use and our setup in Section~\ref{sec:methods}. We show the evolution of the binary system from inspiral to ignition and the subsequent explosion, and follow the ejecta of the explosion until they are in homologous expansion in Section~\ref{sec:explosion}. We compare the properties of the ejecta of both scenarios in Section~\ref{sec:ejecta} and their synthetic observables in Section~\ref{sec:observables}. We discuss our results in a broader context and their implications as well as the next steps in Section~\ref{sec:discussion}. 

\begin{figure*}
\includegraphics[width=0.97\textwidth]{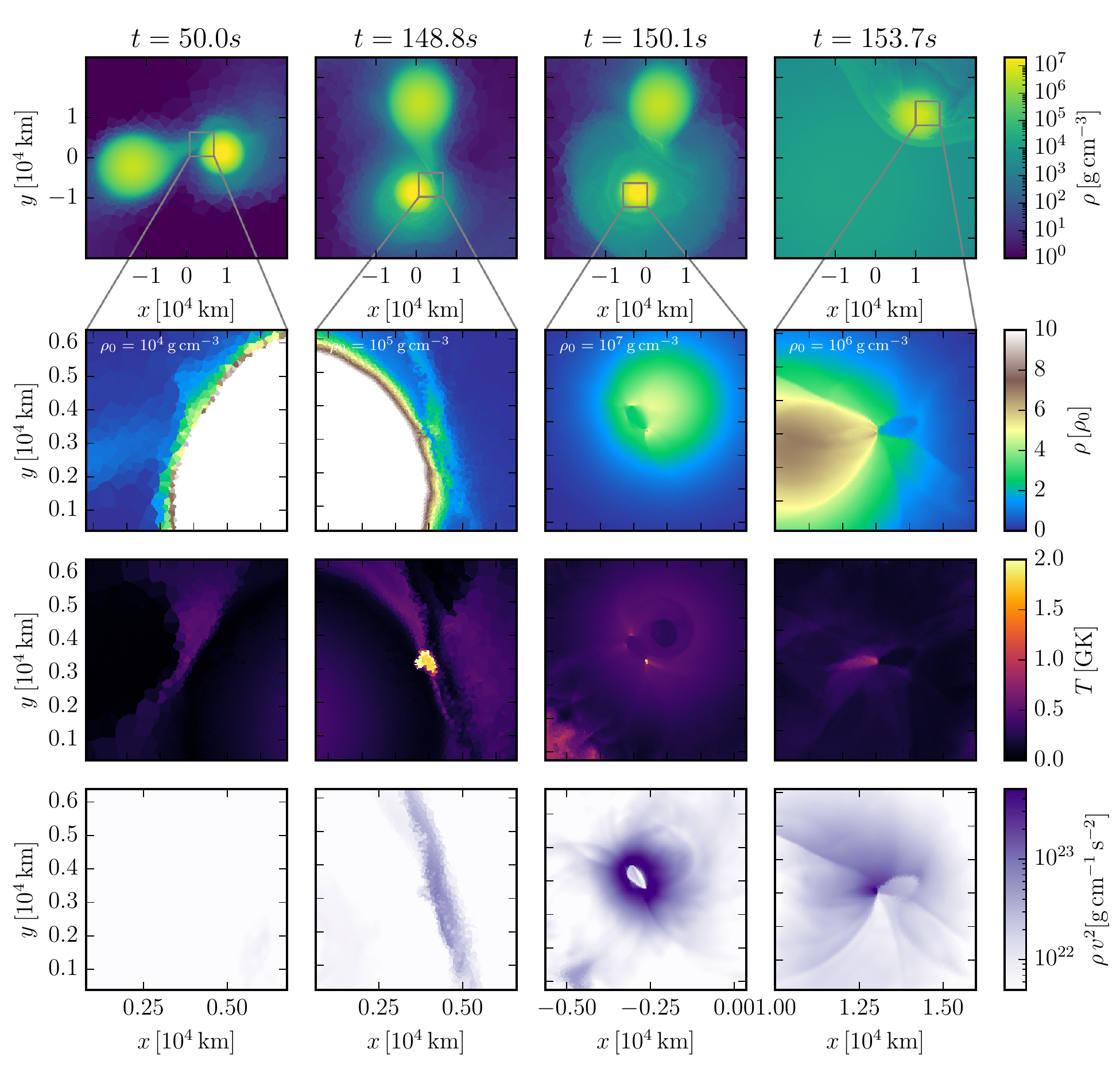}

\caption{Time evolution of the binary system. The columns show the binary when the artificial inspiral phase ends ($t=50\,\mathrm{s}$, first column), when the helium detonation on the primary WD ignites ($t=148.8\,\mathrm{s}$, second column), when the carbon detonation in the primary WD ignites ($t=150.1\,\mathrm{s}$, third column), and when the shock converges in the core of the secondary WD ($t=153.7\,\mathrm{s}$, fourth column). The first row shows a slice of density through the full system. The other rows show slices of density, temperature, and kinetic energy density in a zoomed-in region of interest at each time. For the first column the zoom region is centred on the impact point of the accretion stream onto the primary WD. For the second and third columns it is centred on the ignition points of helium and carbon detonation in the primary WD, respectively. Finally, the zoom-in of the fourth column is centred on the convergence point of the shock in the secondary WD.}
\label{fig:merger}
\end{figure*}

\section{Methods}

\label{sec:methods}

Our simulation pipeline consists of four parts: We first generate one-dimensional (1D) profiles of the two white dwarfs in hydrostatic equilibrium. We then use these 1D profiles to generate 3D white dwarfs in \textsc{arepo} \citep{Arepo,Pakmor2016,Weinberger2020} where we simulate the inspiral and the explosion of the binary system until the ejecta are in homologous expansion. We then postprocess trajectories of Lagrangian tracer particles that we record during the explosion with a $384$ isotope nuclear reaction network to obtain detailed isotopic abundances of the ejecta \citep{Seitenzahl2010,Pakmor2012,Seitenzahl2017}. Finally we use the Monte-Carlo radiative-transfer code \textsc{artis} \citep{Kromer2009,Sim2007} to compute light curves and spectra of the explosion.

We generate 1D profiles of two carbon-oxygen white dwarfs in hydrostatic equilibrium with a constant temperature $T=5\times 10^5\,\mathrm{K}$, masses of $1.05\,\mathrm{M_\odot}$ and $0.7\,\mathrm{M_\odot}$, and central densities of $4.8 \times 10^7\,\mathrm{g\,cm}^{-3}$ and $6.3 \times 10^6\,\mathrm{g\,cm}^{-3}$, respectively. We set the composition to pure helium in the outermost $0.03\,\mathrm{M_\odot}$ and assume a sharp boundary between the helium shell and the carbon-oxygen core. We set the mass fractions of carbon and oxygen in the core to $0.5$.

The composition of both white dwarfs is simplified compared to realistic binary systems of two carbon-oxygen white dwarfs in which the compositions of both white dwarfs will depend on their formation history including potentially several periods of earlier binary interactions \citep{Ruiter2013}. The helium shell masses in our model are likely optimistically large for most double-white-dwarf systems and the consequences of smaller helium shell masses need to be explored in future work.

Note that such massive helium shells may be possible for some systems. Canonical stellar-evolution simulations predict the majority of the helium present on a post-AGB white dwarf progenitor to be ejected. As recently shown \citep[e.g.][]{Zenati2019}, low-mass ($M<1\,\mathrm{M_\odot}$) hydrogen-depleted stars, such as sdB/O stars, naturally evolve into carbon-oxygen white dwarfs with quite massive ($M>0.05\,\mathrm{M_\odot}$) helium envelopes. A system composed of two such objects would naturally account for the presence of the massive helium shells in our simulated system. A system composed of an sdB star and a white dwarf companion will, given the correct orbital separation when it forms, undergo stable mass transfer and thus generate a sufficiently large helium shell on the accreting white dwarf to be compatible with our scenario \citep{Neunteufel2016,Bauer2021,Pelisoli2021}. The sdB star will then naturally evolve into a second carbon-oxygen white dwarf with a substantial helium shell \citep{Zenati2019}, as predicted for the observed system HD\,265435 \citep{Pelisoli2021}.

To simulate the full dynamical evolution of the binary system in 3D, we use the moving-mesh hydrodynamics code \textsc{arepo}. In \textsc{arepo} we follow the dynamical evolution of the binary system from inspiral to explosion. \textsc{arepo} discretises space on a moving Voronoi mesh that is constructed from a set of mesh-generating points. These points each generate a cell and move with the local gas velocity and an additional small velocity correction to keep the mesh regular. This results in an almost Lagrangian evolution of the mesh. The fluid quantities on the mesh are evolved with a second-order finite-volume scheme \citep{Pakmor2016}. Fluxes over interfaces are calculated using the HLLC Riemann solver in the moving frame of the interface \citep{Pakmor2011b}. We employ explicit refinement and de-refinement when the mass of a cell is larger than twice or smaller than half of the target mass resolution (set to $10^{-7}\,\mathrm{M_\odot}$ for all simulations in this paper). Additional refinement is triggered when the volume of a cell is more than $10$ times larger than its smallest direct neighbour to avoid large resolution gradients in the mesh at steep density gradients. Moreover we enforce a maximum volume for cells of $10^{30}\,\mathrm{cm^3}$ to prevent de-refinement of the background mesh.

We use the \textsc{helmholtz} equation of state \citep{Timmes2000} to model the partially degenerate electron-positron gas, ions with Coulomb corrections, and radiation. Moreover, we fully couple a $55$-isotope nuclear reaction network to the simulation \citep{Pakmor2012,Pakmor2021} with the \textsc{jina} reaction rates \citep{Cyburt2010}. The nuclear reaction network is active for all cells with $T>10^6\,\mathrm{K}$. We do not use a burning limiter in the simulations shown in this paper, similar to \citet{Townsley2019}. Since a burning limiter leads to stronger nuclear burning, disabling it can be seen as a conservative approach to ignition. \textsc{arepo} solves self-gravity with a one-sided octree solver. We soften the gravitational force to avoid spurious two-body interactions with a softening length of $2.8$ times the radius of a cell, but force the softening to be at least $10\,\mathrm{km}$.

Our initial setup in \textsc{arepo} closely follows \citet{Pakmor2013} and \citet{Pakmor2021}. We first use the 1D profiles and HEALPix tessellations of the unit sphere to generate 3D meshes of both white dwarfs \citep{Pakmor2012,Ohlmann2017} with roughly cubical cells with a mass close to $10^{-7}\,\mathrm{M_\odot}$. We then relax both white dwarfs individually for $10\,\mathrm{s}$ corresponding to five and two dynamical timescales of the $1.05\,\mathrm{M_\odot}$ white dwarf and the $0.7\,\mathrm{M_\odot}$ white dwarf, respectively, to make sure they are stable and to eliminate spurious noise that we introduced when we generated the initial mesh of the 3D representation of the white dwarfs \citep{Pakmor2012,Ohlmann2017}.

After relaxation, we add both white dwarfs together in a single simulation and put them on a circular co-rotating orbit with an initial orbital period of $T=60\,\mathrm{s}$ and a separation of $a=2.8\times10^9\,\mathrm{cm}$. This separation is large enough that the white dwarfs are essentially undisturbed when they suddenly see the gravitational potential of the other white dwarf. The simulation box has a side length of $10^{12}\,\mathrm{cm}$. We fill the background mesh with a density of $10^{-5}\,\mathrm{g\,cm^{-3}}$ to avoid numerical problems with a vacuum but still only add a negligible amount of mass. We keep the mass resolution of the isolated white dwarfs of $10^{-7}\,\mathrm{M_\odot}$.

For the first $t=50\,\mathrm{s}$, we apply an azimuthal acceleration that mimics the loss of angular momentum via gravitational wave emission to shrink the binary system slowly. To make the simulation feasible we scale it such that the separation decreases at a constant rate $\frac{da}{dt} = 10^2\,\mathrm{km\,s}^{-1}$ \citep{Pakmor2021}. When we stop this force at $t=50\,\mathrm{s}$, the binary orbit has shrunk to $a=2.2\times10^9\,\mathrm{cm}$ and the total angular momentum is now conserved for the rest of the simulation. 

For nucleosynthetis postprocessing we include $10^6$ Lagrangian tracer particles in the \textsc{arepo} simulation. Their initial positions are sampled from the initial mass distribution in the binary system. We record trajectories of position, density, and temperature of those tracer particles with a cadence of $10^{-3}\,\mathrm{s}$. We then postprocess the trajectories with a much larger $384$ isotope reaction network. The initial composition for the postprocessing includes the full isotopic solar composition for $Z_\odot=0.0134$ \citep{Asplund2009} where we assume that all carbon, oxygen, and nitrogen atoms of the solar composition have been converted to $^{22}$Ne during CNO-cycle burning and subsequent helium burning. Including the full solar composition also in unburned material is relevant for the blue part of synthetic spectra \citep{Foley2012}. To normalise the abundances we reduce the mass fractions of helium and oxygen by the total solar metallicity added.

We then map the density of the ejecta at the end of the \textsc{arepo} simulation and the final postprocessed composition of the tracer particles to 1D spherically symmetric profiles with $100$ radial shells. We use these 1D profiles as input to the Monte-Carlo radiative transfer code \textsc{artis} \citep{Kromer2009, Sim2007} and compute synthetic light curves and spectra until $70\,\mathrm{d}$ after the explosion. We use $10^7$ Monte Carlo packets.

\begin{figure*}
\includegraphics[width=0.97\textwidth]{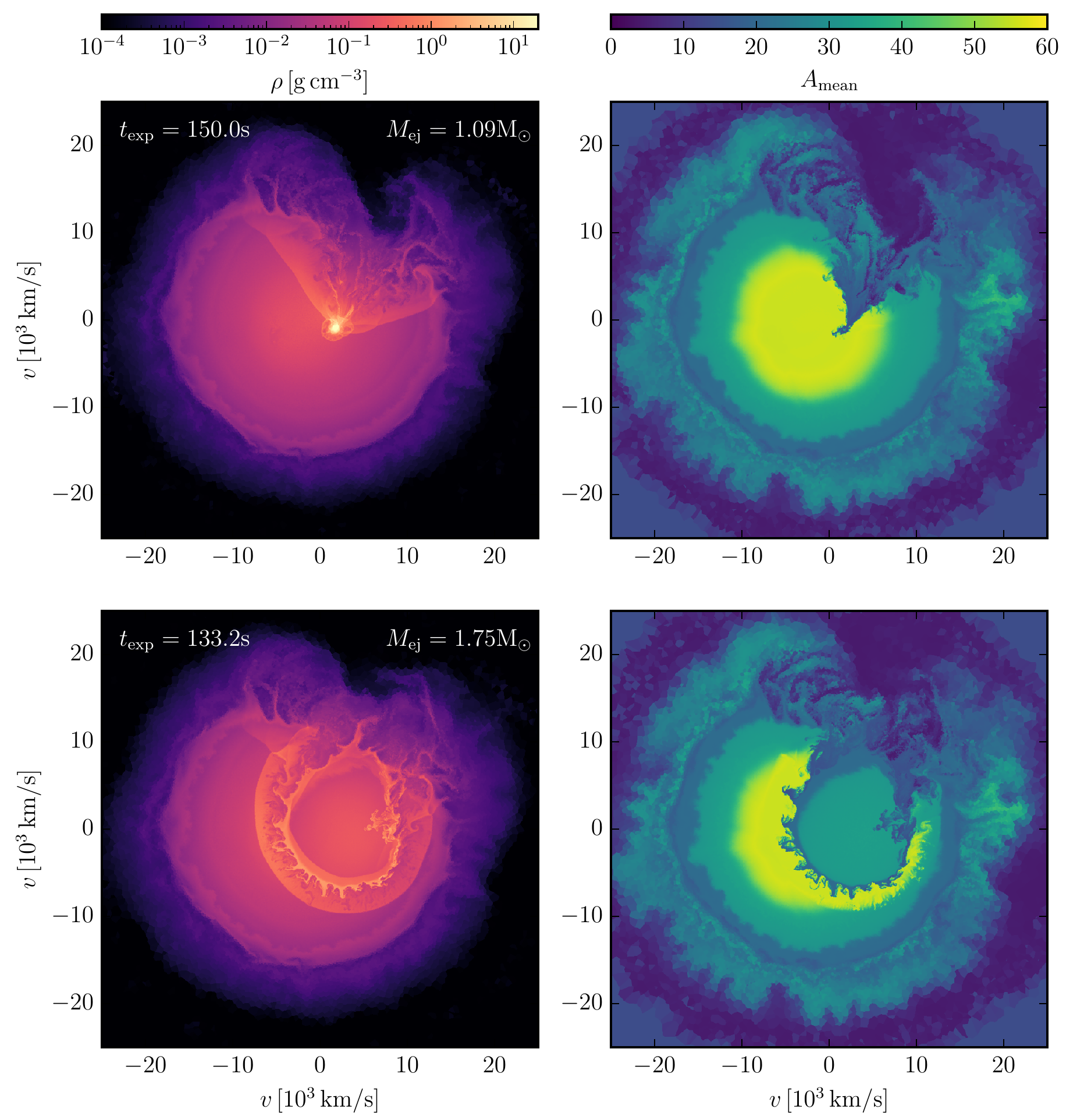}

\caption{Density (left colum) and mean atomic weight (right column) in a slice in the $x$--$y$ plane. The top row shows the ejecta of the primary white dwarf $150\,\mathrm{s}$ after the ignition of the carbon detonation in the primary white dwarf if the secondary white dwarf does not explode. The bottom row shows the total ejecta $133\,\mathrm{s}$ after the ignition of the carbon detonation in the primary white dwarf if the secondary white dwarf also detonates. At these times both ejecta are fully in homologous expansion. The explosion of the secondary white dwarf drastically changes the inner ejecta, but leaves the outer ejecta of the explosion of the primary white dwarf unchanged.}
\label{fig:ejecta}
\end{figure*}

\section{Inspiral and explosion}

\label{sec:explosion}

\begin{table*}
\centering
\begin{tabular}{ c | c | r r | r r r r }
  & Unit & OneExpl & TwoExpl & prim. He det & prim. CO det & sec. He det & sec. CO det\\
  \hline
  $\Delta \mathrm{E_{nuc}}$ & $\mathrm{erg}$ & $1.4 \times 10^{51}$ & $1.9 \times 10^{51}$ & $8.3 \times 10^{49}$ & $1.4 \times 10^{51}$ & $6.7 \times 10^{49}$ & $5.6 \times 10^{50}$ \\
  $M_\mathrm{ej}$ & $\mathrm{M_\odot}$ & $1.09$ & $1.75$ &  $7.8\times 10^{-2}$ & $0.99$ & $1.5\times 10^{-2}$ & $0.56$ \\ 
  $t_\mathrm{start}$ & $\mathrm{s}$ & & & $148.8$ & $150.1$ & $151.0$ & $153.7$ \\
  \hline
   He & $\mathrm{M_\odot}$ & $-1.4 \times 10^{-2}$ & $-1.7 \times 10^{-2}$ & $-1.4 \times 10^{-2}$ & $ 9.4 \times 10^{-4}$ & $-2.2 \times 10^{-3}$ & $-1.8 \times 10^{-3}$ \\
    C & $\mathrm{M_\odot}$ & $-5.1 \times 10^{-1}$ & $-7.8 \times 10^{-1}$ & $-1.7 \times 10^{-2}$ & $-4.9 \times 10^{-1}$ & $-8.0 \times 10^{-4}$ & $-2.7 \times 10^{-1}$ \\
    O & $\mathrm{M_\odot}$ & $-4.1 \times 10^{-1}$ & $-5.0 \times 10^{-1}$ & $-5.0 \times 10^{-3}$ & $-4.0 \times 10^{-1}$ & $-1.3 \times 10^{-3}$ & $-9.0 \times 10^{-2}$ \\
   Ne & $\mathrm{M_\odot}$ & $-6.3 \times 10^{-3}$ & $-2.4 \times 10^{-3}$ & $ 5.6 \times 10^{-3}$ & $-1.1 \times 10^{-2}$ & $ 3.4 \times 10^{-4}$ & $ 3.1 \times 10^{-3}$ \\
   Na & $\mathrm{M_\odot}$ & $ 2.5 \times 10^{-5}$ & $ 1.5 \times 10^{-4}$ & $ 4.2 \times 10^{-5}$ & $-1.4 \times 10^{-6}$ & $ 4.6 \times 10^{-6}$ & $ 1.0 \times 10^{-4}$ \\
   Mg & $\mathrm{M_\odot}$ & $ 1.2 \times 10^{-2}$ & $ 3.4 \times 10^{-2}$ & $ 5.8 \times 10^{-3}$ & $ 5.9 \times 10^{-3}$ & $ 6.0 \times 10^{-4}$ & $ 2.2 \times 10^{-2}$ \\
   Al & $\mathrm{M_\odot}$ & $ 4.3 \times 10^{-4}$ & $ 1.4 \times 10^{-3}$ & $ 2.8 \times 10^{-4}$ & $ 1.6 \times 10^{-4}$ & $ 2.7 \times 10^{-5}$ & $ 9.1 \times 10^{-4}$ \\
   Si & $\mathrm{M_\odot}$ & $ 2.3 \times 10^{-1}$ & $ 4.4 \times 10^{-1}$ & $ 8.8 \times 10^{-3}$ & $ 2.2 \times 10^{-1}$ & $ 1.1 \times 10^{-3}$ & $ 2.0 \times 10^{-1}$ \\
    P & $\mathrm{M_\odot}$ & $ 4.7 \times 10^{-4}$ & $ 1.3 \times 10^{-3}$ & $ 1.8 \times 10^{-4}$ & $ 2.8 \times 10^{-4}$ & $ 3.1 \times 10^{-5}$ & $ 7.6 \times 10^{-4}$ \\
    S & $\mathrm{M_\odot}$ & $ 1.4 \times 10^{-1}$ & $ 2.3 \times 10^{-1}$ & $ 3.8 \times 10^{-3}$ & $ 1.3 \times 10^{-1}$ & $ 1.0 \times 10^{-3}$ & $ 9.7 \times 10^{-2}$ \\
   Cl & $\mathrm{M_\odot}$ & $ 2.6 \times 10^{-4}$ & $ 4.2 \times 10^{-4}$ & $ 1.4 \times 10^{-4}$ & $ 7.0 \times 10^{-5}$ & $ 5.3 \times 10^{-5}$ & $ 1.6 \times 10^{-4}$ \\
   Ar & $\mathrm{M_\odot}$ & $ 2.5 \times 10^{-2}$ & $ 3.9 \times 10^{-2}$ & $ 1.3 \times 10^{-3}$ & $ 2.4 \times 10^{-2}$ & $ 5.0 \times 10^{-4}$ & $ 1.4 \times 10^{-2}$ \\
    K & $\mathrm{M_\odot}$ & $ 4.7 \times 10^{-4}$ & $ 5.5 \times 10^{-4}$ & $ 3.2 \times 10^{-4}$ & $ 5.6 \times 10^{-5}$ & $ 1.1 \times 10^{-4}$ & $ 6.6 \times 10^{-5}$ \\
   Ca & $\mathrm{M_\odot}$ & $ 2.4 \times 10^{-2}$ & $ 3.2 \times 10^{-2}$ & $ 3.7 \times 10^{-3}$ & $ 2.0 \times 10^{-2}$ & $ 3.7 \times 10^{-4}$ & $ 7.8 \times 10^{-3}$ \\
   Sc & $\mathrm{M_\odot}$ & $ 1.0 \times 10^{-4}$ & $ 1.1 \times 10^{-4}$ & $ 8.6 \times 10^{-5}$ & $ 1.2 \times 10^{-6}$ & $ 1.8 \times 10^{-5}$ & $ 7.6 \times 10^{-7}$ \\
   Ti & $\mathrm{M_\odot}$ & $ 1.6 \times 10^{-3}$ & $ 1.6 \times 10^{-3}$ & $ 1.5 \times 10^{-3}$ & $ 3.4 \times 10^{-5}$ & $ 3.4 \times 10^{-5}$ & $ 2.9 \times 10^{-5}$ \\
    V & $\mathrm{M_\odot}$ & $ 1.6 \times 10^{-4}$ & $ 1.6 \times 10^{-4}$ & $ 1.6 \times 10^{-4}$ & $ 1.5 \times 10^{-6}$ & $ 3.3 \times 10^{-6}$ & $ 1.9 \times 10^{-6}$ \\
   Cr & $\mathrm{M_\odot}$ & $ 2.3 \times 10^{-3}$ & $ 2.5 \times 10^{-3}$ & $ 1.8 \times 10^{-3}$ & $ 5.2 \times 10^{-4}$ & $ 7.3 \times 10^{-6}$ & $ 1.5 \times 10^{-4}$ \\
   Mn & $\mathrm{M_\odot}$ & $ 2.8 \times 10^{-4}$ & $ 3.1 \times 10^{-4}$ & $ 2.1 \times 10^{-4}$ & $ 8.2 \times 10^{-5}$ & $ 5.2 \times 10^{-7}$ & $ 1.5 \times 10^{-5}$ \\
   Fe & $\mathrm{M_\odot}$ & $ 2.3 \times 10^{-2}$ & $ 3.0 \times 10^{-2}$ & $ 1.1 \times 10^{-3}$ & $ 2.2 \times 10^{-2}$ & $-2.7 \times 10^{-6}$ & $ 6.5 \times 10^{-3}$ \\
   Co & $\mathrm{M_\odot}$ & $ 3.0 \times 10^{-3}$ & $ 3.4 \times 10^{-3}$ & $ 1.6 \times 10^{-4}$ & $ 2.9 \times 10^{-3}$ & $ 5.7 \times 10^{-6}$ & $ 3.4 \times 10^{-4}$ \\
   Ni & $\mathrm{M_\odot}$ & $ 4.7 \times 10^{-1}$ & $ 4.7 \times 10^{-1}$ & $ 7.1 \times 10^{-4}$ & $ 4.7 \times 10^{-1}$ & $ 8.6 \times 10^{-6}$ & $ 4.5 \times 10^{-3}$ \\
   Cu & $\mathrm{M_\odot}$ & $ 6.0 \times 10^{-4}$ & $ 1.1 \times 10^{-3}$ & $ 1.7 \times 10^{-5}$ & $ 1.0 \times 10^{-3}$ & $ 1.5 \times 10^{-6}$ & $ 1.8 \times 10^{-5}$ \\
   Zn & $\mathrm{M_\odot}$ & $ 6.8 \times 10^{-3}$ & $ 6.4 \times 10^{-3}$ & $ 3.8 \times 10^{-5}$ & $ 6.3 \times 10^{-3}$ & $ 1.4 \times 10^{-6}$ & $ 1.0 \times 10^{-4}$ \\
   \hline
   $^{44}$Ti & $\mathrm{M_\odot}$ & $ 1.5 \times 10^{-3}$ & $ 1.6 \times 10^{-3}$ & $ 1.5 \times 10^{-3}$ & $ 3.1 \times 10^{-5}$ & $ 3.0 \times 10^{-5}$ & $ 2.2 \times 10^{-5}$ \\
   $^{48}$Cr & $\mathrm{M_\odot}$ & $ 2.0 \times 10^{-3}$ & $ 2.1 \times 10^{-3}$ & $ 1.7 \times 10^{-3}$ & $ 3.6 \times 10^{-4}$ & $ 6.6 \times 10^{-6}$ & $ 5.5 \times 10^{-5}$ \\
   $^{52}$Fe & $\mathrm{M_\odot}$ & $ 8.2 \times 10^{-3}$ & $ 8.6 \times 10^{-3}$ & $ 1.0 \times 10^{-3}$ & $ 7.2 \times 10^{-3}$ & $ 5.2 \times 10^{-6}$ & $ 2.9 \times 10^{-4}$ \\
   $^{55}$Co & $\mathrm{M_\odot}$ & $ 3.0 \times 10^{-3}$ & $ 3.3 \times 10^{-3}$ & $ 1.4 \times 10^{-4}$ & $ 2.9 \times 10^{-3}$ & $ 4.8 \times 10^{-7}$ & $ 3.2 \times 10^{-4}$ \\
   $^{56}$Ni & $\mathrm{M_\odot}$ & $ 4.5 \times 10^{-1}$ & $ 4.6 \times 10^{-1}$ & $ 5.5 \times 10^{-4}$ & $ 4.5 \times 10^{-1}$ & $ 4.5 \times 10^{-6}$ & $ 3.7 \times 10^{-3}$ \\
   $^{57}$Ni & $\mathrm{M_\odot}$ & $ 8.2 \times 10^{-3}$ & $ 8.3 \times 10^{-3}$ & $ 7.6 \times 10^{-5}$ & $ 8.1 \times 10^{-3}$ & $ 5.5 \times 10^{-7}$ & $ 1.3 \times 10^{-4}$ \\
   $^{58}$Ni & $\mathrm{M_\odot}$ & $ 8.8 \times 10^{-3}$ & $ 9.2 \times 10^{-3}$ & $ 4.6 \times 10^{-5}$ & $ 8.7 \times 10^{-3}$ & $-6.9 \times 10^{-7}$ & $ 4.5 \times 10^{-4}$ \\
\end{tabular}
\caption{Energy release, mass, and mass change of all elements up to Zn and selected isotopes for the total ejecta of both simulations and split into the helium and carbon detonations of the primary and secondary white dwarf. The first three detonations are the same for both explosion models, the secondary carbon detonation only occurs in the model that we restart from the time when the shockwave converges in the core of the secondary white dwarf. The energy release and abundances are computed from the postprocessed tracer particles at the end of the simulation.}
\label{tab:nuclearburning}
\end{table*}

We show the binary system at different times in Figure~\ref{fig:merger}: at the time when we stop the inspiral, when the helium detonation ignites on the surface of the primary white dwarf, when the carbon detonation ignites in the core of the primary white dwarf, and when the shock converges in the core of the secondary white dwarf.

The primary white dwarf undergoes the classic double-detonation mechanism. The helium detonation ignites at $t=148.8\,\mathrm{s}$. At this time the binary system has lost $10^{-4}\,\mathrm{M_\odot}$ of $^4\mathrm{He}$. Roche-lobe overflow has transferred $10^{-2}\,\mathrm{M_\odot}$ of $^4\mathrm{He}$ from the secondary white dwarf to the primary white dwarf. The helium detonation ignites close to the point where the accretion stream hits the surface of the primary white dwarf (see second column of Figure~\ref{fig:merger}).

The helium detonation then wraps around the primary white dwarf burning its helium shell and sending a shockwave into its core. This shockwave converges in a single point in the core of the primary white dwarf and ignites a carbon detonation there at $t=150.1\,\mathrm{s}$ (see third column of Figure~\ref{fig:merger}) $25\,\mathrm{km}$ from its centre. The carbon detonation completely burns and destroys the primary white dwarf.

The double-detonation mechanism initially repeats on the secondary white dwarf: The shockwave of the explosion of the primary white dwarf hits the secondary white dwarf. It ignites the helium shell of the secondary white dwarf at around $t=151\,\mathrm{s}$ and additionally sends a shockwave into its core directly. This shockwave, supported by the helium detonation burning the remaining helium around the secondary white dwarf converges in the core of the secondary white dwarf at $t=153.7\,\mathrm{s}$ (see fourth column of Figure~\ref{fig:merger}).

In contrast to the shock in the primary white dwarf, the converging shockwave in the secondary white dwarf fails to ignite a carbon detonation in its core in our simulation. Instead the secondary white dwarf in our simulation survives until the end of the simulation $100\,\mathrm{s}$ later. We call this model that destroys only the primary white dwarf "OneExpl".

The physical conditions at the point where the shock converges are a density of $\rho=5.4\times10^6\,\mathrm{g\,cm^{-3}}$ and a temperature of $T=1.1\times 10^9\,\mathrm{K}$. These conditions are possibly sufficient to ignite a carbon detonation \citep{Seitenzahl2009}. However, the lower density compared to the convergence point in the primary white dwarf at $\rho=2\times10^7\,\mathrm{g\,cm^{-3}}$, where a carbon detonation ignites successfully, requires significantly higher numerical resolution than achieved in our simulation \citep{Seitenzahl2009}.

We therefore argue that a detonation initiation in the secondary is physically plausible and restart our simulation at $t=153.7\,\mathrm{s}$, this time igniting a carbon detonation by hand at the convergence point of the shockwave. We increase the temperature to $T_\mathrm{new}=5\times10^9\,\mathrm{K}$ in a sphere of radius $r=10^2\,\mathrm{km}$ around the convergence point. This injects $4.9\times 10^{46}\,\mathrm{erg}$ into $113$ cells with a total mass of $10^{-5}\,\mathrm{M_\odot}$. In this simulation, the second carbon detonation burns the secondary white dwarf completely and destroys it as well. We again continue the simulation for $80\,\mathrm{s}$ to make sure the ejecta are in homologous expansion at the end of the simulation. We call this model that destroys both white dwarfs "TwoExpl".

To understand the energetics and the timing of the detonations, we quantify the time for which a detonation is active as the difference between the time when it has released $1\%$ of its total nuclear energy and the time when it has released $99\%$ of it. A detailed table of the energy release and changes to the nuclear composition by the four detonations is shown in Table~\ref{tab:nuclearburning}.

For this analysis, we associate all tracer particles that change their total nuclear binding energy by at least $10$ per cent with the different detonations in the simulation. We associate all tracer particles that burn before the carbon detonation in the primary white dwarf ignites as part of the helium detonation on the primary white dwarf. From the remaining tracer particles we associate those that burn before the secondary carbon detonation ignites and that have a helium mass fraction less than $0.1$ in their initial composition with the carbon detonation of the primary white dwarf. We associate similar tracer particles with helium with the helium detonation of the secondary white dwarf. Finally, we associate all tracer particles that burn but are not associated with all the other three detonations with the carbon detonation in the secondary white dwarf. Note that that this category remains empty in the OneExpl simulation.

The helium detonation on the surface of the primary white dwarf releases $8.3\times 10^{49}\,\mathrm{erg}$ and burns for $0.9\,\mathrm{s}$. The carbon detonation in the core of the primary white dwarf ignites $1.3\,\mathrm{s}$ after the ignition of the helium detonation. This time lag is typical for medium-strength helium detonations in the double-detonation mechanism \citep{Fink2010}.

The carbon detonation in the primary white dwarf burns for $0.4\,\mathrm{s}$ and releases $1.4\times 10^{51}\,\mathrm{erg}$. Its explosion ashes then have some time to expand before the double-detonation mechanism starts for the secondary white dwarf. The helium burning around the secondary white dwarf releases another $6.7\times 10^{49}\,\mathrm{erg}$.

The shockwave converges in the core of the secondary white dwarf $3.3\,\mathrm{s}$ after the primary white dwarf finished burning and started to expand. At this point almost all of the ejecta of the primary white dwarf have already passed beyond the secondary white dwarf, which is now essentially at the centre of the ejecta of the primary white dwarf.
For our restarted simulation, in which we ignite the carbon detonation in the secondary white dwarf by hand, the secondary white dwarf burns completely in $0.7\,\mathrm{s}$. It releases $5.6\times 10^{50}\,\mathrm{erg}$. The expansion of the ejecta of the secondary white dwarf starts $4.0\,\mathrm{s}$ after the expansion of the ejecta of the primary white dwarf.

The ejecta of the secondary white dwarf expand into the centre of the now low-density ejecta of the primary white dwarf. This configuration is very different from violent merger models \citep{Pakmor2012b}, where primary and secondary white dwarf explode roughly at the same time. For those models, there is only a difference of $\Delta t < 0.5\,\mathrm{s}$ between the explosion of the primary white dwarf and the explosion of the partially disrupted secondary white dwarf, which is set by the travel time of the carbon detonation from the primary to the secondary white dwarf.

\begin{figure}
\includegraphics[width=0.97\linewidth]{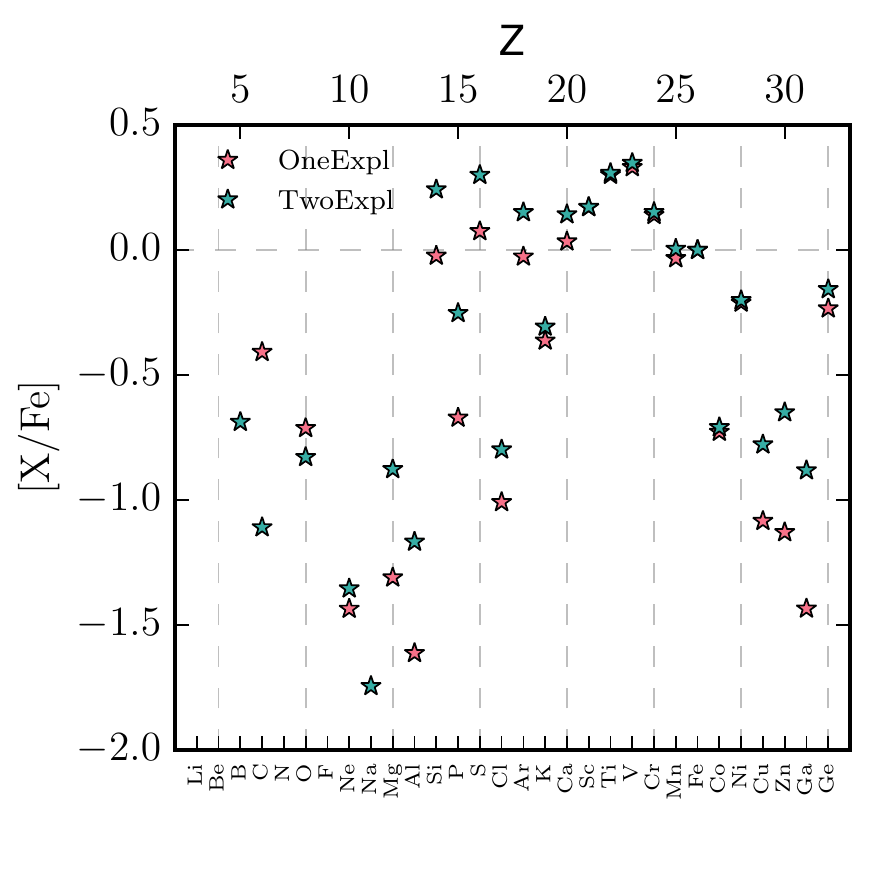}

\caption{Logarithm of the elemental composition relative to iron of the ejecta of both models $10^6\,\mathrm{yr}$ after the explosion relative to the solar composition \citep{Asplund2009}, i.e. taking into account radioactive decay of unstable isotopes in the ejecta. The TwoExpl model has slightly supersolar yields for intermediate mass elements and solar yields for manganese.}
\label{fig:yields}
\end{figure}

\begin{figure*}
\includegraphics[width=0.97\textwidth]{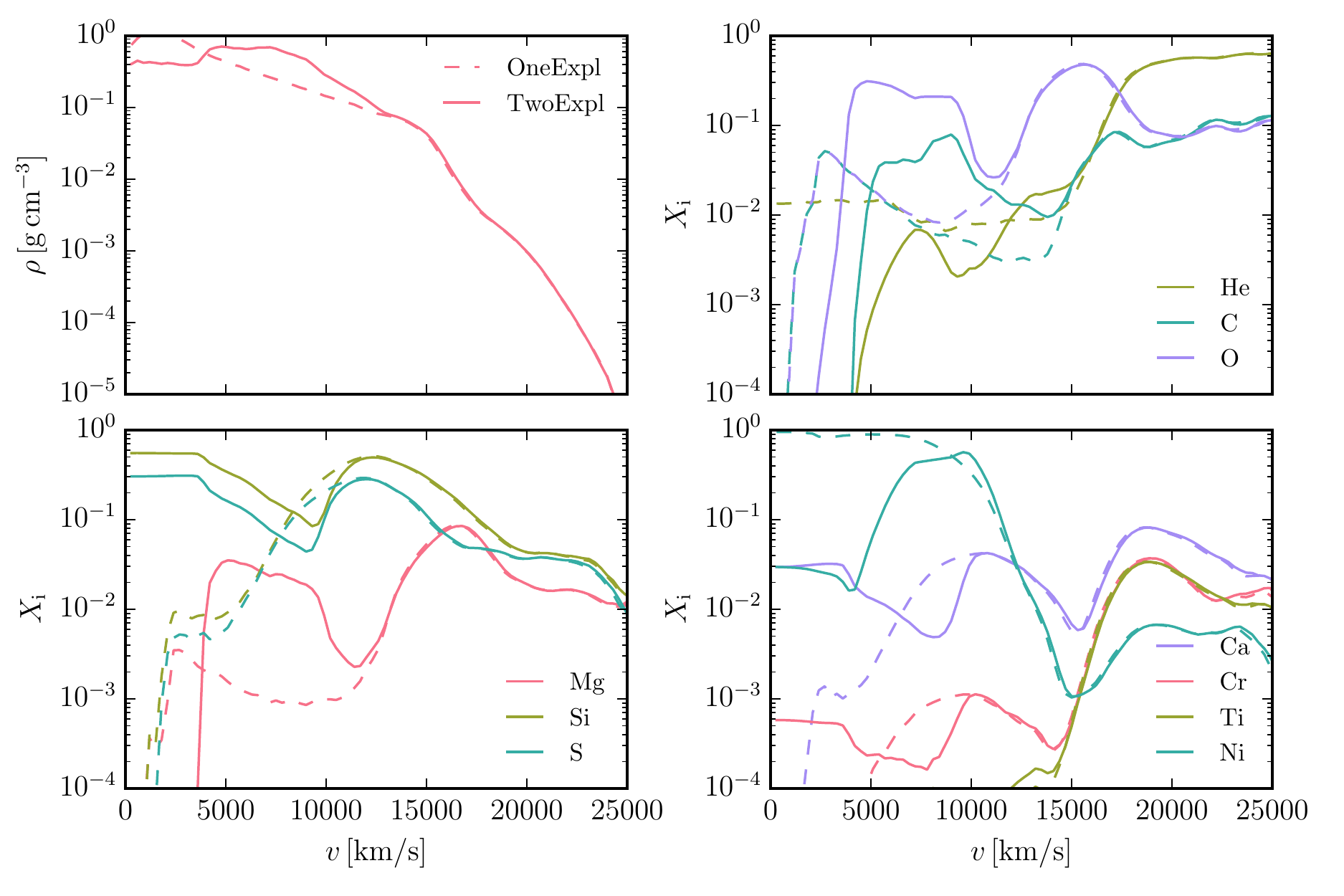}

\caption{Radial profiles of the mass density and mass fraction of various elements of the unbound ejecta $100\,\mathrm{s}$ after the explosion when the ejecta are fully in homologous expansion. Solid lines show the profiles of the model in which both white dwarfs explode, dashed lines show the model in which the secondary white dwarf survives. For $v\gtrsim 15\,000\,\mathrm{km\,s}^{-1}$ the ejecta of both models have identical density and composition.}
\label{fig:ejecta_profiles}
\end{figure*}

\section{Ejecta structure and composition}
\label{sec:ejecta}

\begin{figure}
\includegraphics[width=0.97\linewidth]{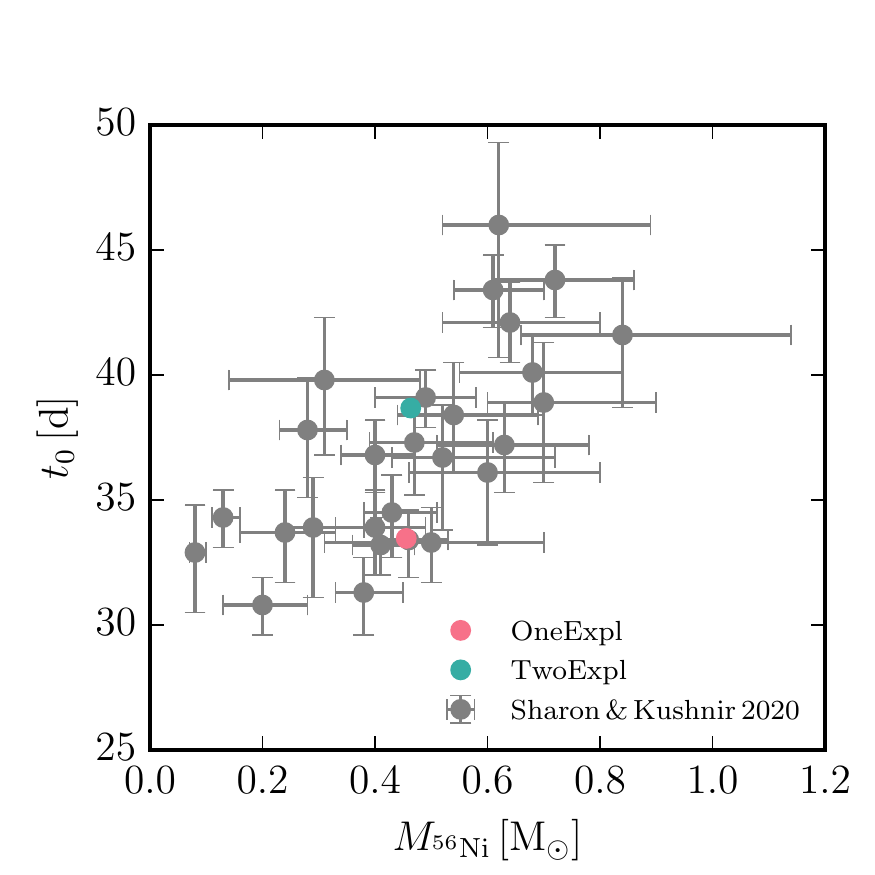}

\caption{Gamma-ray escape time $t_0$ versus the mass of $\mathrm{^{56}Ni}$ in the ejecta for both explosion models and a set of SNe~Ia with well-observed bolometric light curves \citep{Sharon2020}. The TwoExpl model has a significantly longer gamma-ray escape time than the OneExpl model.}
\label{fig:ejecta_t0}
\end{figure}

We show slices of the ejecta of both explosion models in homologous expansion in the orbital plane of the initial binary system in Figure~\ref{fig:ejecta}. The top row shows the ejecta of the simulation in which the secondary white dwarf survives, which we will call "OneExpl". The bottom row shows the ejecta of the simulation in which we ignited the carbon detonation in the secondary white dwarf by hand and where the secondary explodes as well, which we name "TwoExpl". The left column shows the density of the ejecta, the right column their mean atomic weight.

\subsection{Global structure of the ejecta}

We immediately notice that the outer ejecta of both explosions are very similar. Focusing on the TwoExpl simulation we see that the ejecta of the secondary white dwarf expand into the ejecta of the primary white dwarf and compress their inner parts. They stall around $10\,000\,\mathrm{km\,s}^{-1}$ depending on the orientation angle relative to the position of primary and secondary white dwarf at the time of explosion and compress the ejecta of the primary white dwarf up to velocities of about $15\,000\,\mathrm{km\,s}^{-1}$.

The obvious global feature of the ejecta is the low density cone in the ejecta of the primary white dwarf in the OneExpl model where the secondary white dwarf blocks them. In the TwoExpl model this cone is filled by the ejecta of the secondary white dwarf. As a result the density distribution of the TwoExpl model is more symmetric than the density distribution of the OneExpl model. However, the TwoExpl model has a very different composition in the direction of the now-filled cone. This cone is most pronounced in the plane of rotation of the binary system shown in Figure~\ref{fig:ejecta}, but will likely only affect a small fraction of the viewing angles to the explosion.

Apart from the cone the outer ejecta are close to spherical symmetry, similar to idealised models of explosions of isolated sub-Chandrasekhar-mass white dwarfs. The symmetry of the density distribution of the outer ejecta directly translates to the continuum polarisation of the explosion, so we expect only a weak signal. Line polarisation that is sensitive to asymmetries in the composition could be significantly stronger because the mean atomic weight distribution is slightly offset at large velocities and more asymmetric in the inner parts of both simulations. 

\subsection{Ejecta composition}

We collect the detailed yields of all elements up to Zn and selected isotopes in Table~\ref{tab:nuclearburning}. The energy release from the helium detonations is about $5\%$ of the energy release of the carbon detonations. Thus the latter completely dominate the energetics of the explosion.

An important difference between the two models is the total ejecta mass. The ejecta of the OneExpl model contain $1.09\,\mathrm{M_\odot}$ - the initial mass of the primary white dwarf and a little material from the secondary white dwarf that was accreted onto the primary before the explosion, or material that was stripped from the secondary white dwarf by the ejecta of the primary white dwarf. In contrast, the TwoExpl model does not leave any bound remnant behind and the ejecta contain the full $1.75\,\mathrm{M_\odot}$ of the initial binary system, significantly more than the mass of a Chandrasekhar-mass carbon-oxygen white dwarf of $1.38\,\mathrm{M_\odot}$.

The helium ashes on the primary white dwarf are dominated by intermediate-mass elements, i.e. silicon, sulfur, and calcium. They also contain small amounts of titanium ($1.5\times 10^{-3}\,\mathrm{M_\odot}$) and chromium ($1.8\times10^{-3}\,\mathrm{M_\odot}$) that are irrelevant for the energetics but important for the synthetic observables (see also Section~\ref{sec:observables}).

Essentially all radioactive $^{56}\mathrm{Ni}$ is produced from the burning of the carbon-oxygen core of the primary white dwarf. Its burning products are consistent with classic idealised sub-Chandrasekhar-mass detonation models \citep{Sim2010, Shen2018} and are dominated by $^{56}\mathrm{Ni}$ and intermediate-mass elements.

The helium shell of the secondary white dwarf is burned in both simulations, independently of the final fate of the secondary white dwarf. It burns a total of $1.5\times 10^{-2}\,\mathrm{M_\odot}$ of helium mixed with some carbon and oxygen into intermediate-mass elements. It does not produce any titanium or chromium.

Even in the TwoExpl simulation, in which the carbon-oxygen core of the secondary white dwarf is fully burned, this burning does not produce any relevant amount of iron-group elements due to the low central density of the secondary white dwarf. Instead the ashes of the secondary white dwarf consist of intermediate-mass elements dominated by silicon, sulfur, argon, and calcium.

We show the element-wise integrated yields relative to iron of both explosion models relative to the solar composition in Figure~\ref{fig:yields}. Here we assume all isotopes with a halflife shorter than $10^6\,\mathrm{yr}$ have fully decayed. The OneExpl model has about solar yields of silicon, sulfur, argon, and calcium. Those yields are slightly super-solar for the TwoExpl model because the carbon detonation of the secondary white dwarf produces some amount of those elements, but no additional iron. The TwoExpl model produces about $10$ per cent more manganese than the OneExpl model, pushing its manganese yield just above solar, but still below yields predicted for most Chandrasekhar-mass explosions \citep{Seitenzahl2013}. Note, however, that the yields of one explosion simulation can not easily be extrapolated to the integrated yields expected if all normal Type Ia supernovae originate from this scenario because the yields change too much with the brightness of the explosion \citep{Sim2010, Seitenzahl2013b,Gronow2021}.

Note that our explosions produce about a factor of $10$ more $^{44}$Ti than found in Tycho's remnant \citep{Troja2014}. In both of our models almost all $^{44}$Ti is produced in the helium detonation of the primary white dwarf. This discrepancy can thus be alleviated through thinner helium shells \citep{Boos2021}. However, we will need parameter studies of the full system that vary the mass of the helium shell on the primary white dwarf to figure out whether double-detonation models can match this limit and still explode. A direct detection of other radioactive isotopes from supernova remnants could also add important constraints \citep{Panther2021}.

\subsection{Ejecta profiles}

We show 1D spherically averaged symmetric radial profiles of the ejecta of both models for density and mass fractions of the most important elements in Figure~\ref{fig:ejecta_profiles}. We see that the outer parts of both explosions are near identical in density and composition for $v\,\gtrsim\,15\,000\,\mathrm{km\,s}^{-1}$. As we showed in the slice plots before, the inner parts, in contrast, are very different. Thus we expect both explosions to look very similar until some time after maximum brightness when the effective photosphere recedes into the parts of the ejecta that are different. Moreover, we expect nebular spectra of the two explosions that directly probe the inner parts of the ejecta to look very different. However, owing to the complicated structure of the inner ejecta shown in Figure~\ref{fig:ejecta} any faithful synthetic nebular spectra will need to be computed in 3D to take this structure fully into account. We plan to present synthetic 3D nebular spectra of our models in the near future. 

\subsection{Gamma-ray escape time}

After maximum brightness, but before the ejecta become fully optically thin to gamma rays in the nebular phase, we expect differences in the bolometric light curves because the typical density of the $^{56}\mathrm{Ni}$-rich material is much higher in the TwoExpl model than in the OneExpl model (see also Figure~\ref{fig:ejecta}). Figure~\ref{fig:ejecta_t0} shows the gamma-ray escape time $t_0$ which approximates the long-term evolution of the bolometric light curve. Here we compute $t_0$ from the spherically averaged radial density and $^{56}\mathrm{Ni}$ abundance profiles assuming a single scattering event for each $\gamma$ photon before it escapes \citep{Wygoda2019}. We also show fits to $t_0$ from bolometric light curves of a set of well observed normal Type Ia supernovae \citep{Sharon2020}. We can clearly see a difference between both explosion models. The OneExpl model ($t_0=33.5\,\mathrm{d}$, $M_\mathrm{^{56}Ni}=0.456\,\mathrm{M_\odot}$) is very similar to isolated double-detonation simulations that tend to have gamma-ray escape times smaller then observed \citep{Kushnir2020}. The TwoExpl model in which the $^{56}\mathrm{Ni}$-rich material is compressed by the ejecta of the secondary white dwarf has a significantly longer gamma-ray escape time ($t_0=38.7\,\mathrm{d}$, $M_\mathrm{^{56}Ni}=0.463\,\mathrm{M_\odot}$). It is consistent with the largest values found for observed supernovae with the same $^{56}\mathrm{Ni}$ mass. It seems plausible that values between the two extremes could stem from different viewing angles. We aim to test this in the future with full 3D synthetic light curves. Therefore modelling the full double-degenerate binary system rather than models of isolated white dwarfs might be essential to assess the validity of this scenario.

\subsection{Velocity shifts}

A last important property of the ejecta is their global velocity shift relative to the rest frame of the binary system. This directly leads to velocity shifts that can be observed in the nebular phase and have so far mostly been associated with an off-centre ignition in Chandrasekhar-mass scenarios \citep[see, e.g.][]{Maeda2010}. In the pure detonation models discussed here, however, the velocity shifts introduced from an off-centre ignition are likely subdominant compared to the global velocity shifts that originate from the orbital velocity of the exploding white dwarf in the close binary system. In the OneExpl model, where the secondary white dwarf survives, the unbound ejecta move with a velocity of $1100\,\mathrm{km\,s}^{-1}$ [with a velocity vector $\textbf{v} = (-970,\,520,\,0)\,\mathrm{km\,s}^{-1}$]. The centre of mass of the $^{56}\mathrm{Ni}$ moves even faster, with a velocity of $1610\,\mathrm{km\,s}^{-1}$, in a similar but not identical direction [$\textbf{v} = (-1600,\,170,\,25)\,\mathrm{km\,s}^{-1}$], in contrast to off-centre ignition models in which the centre-of-mass velocity of $^{56}\mathrm{Ni}$ is compensated by the centre-of-mass velocity of intermediate-mass elements. The surviving secondary white dwarf moves into the opposite direction of the original ejecta with a velocity of $1790\,\mathrm{km\,s}^{-1}$ and a velocity vector $\textbf{v} = (1580,\,-850,\,0)\,\mathrm{km\,s}^{-1}$.

Obviously, the situation changes in the TwoExpl model where the secondary white dwarf is disrupted as well and no bound remnant remains. The total centre-of-mass velocity of the ejecta is now only $4\,\mathrm{km\,s}^{-1}$. Its deviation from zero is a result of small accumulated inaccuracies in our gravity solver. Since the ejecta of the secondary white dwarf interact strongly with the $^{56}\mathrm{Ni}$-rich material in the centre of the ejecta of the primary white dwarf, the centre-of-mass velocity of the $^{56}\mathrm{Ni}$ increases to $2390\,\mathrm{km\,s}^{-1}$ with a velocity vector $\textbf{v} = (-2300,\,-670,\,30)\,\mathrm{km\,s}^{-1}$, at the very end of the distribution of expected orbital velocities in binary systems of two carbon-oxygen white dwarfs \citep{Shen2018b}. Therefore, looking for velocity shifts that large in nebular spectra of normal Type Ia supernovae may be another way to test this scenario.

\begin{figure*}
\includegraphics[width=0.97\textwidth]{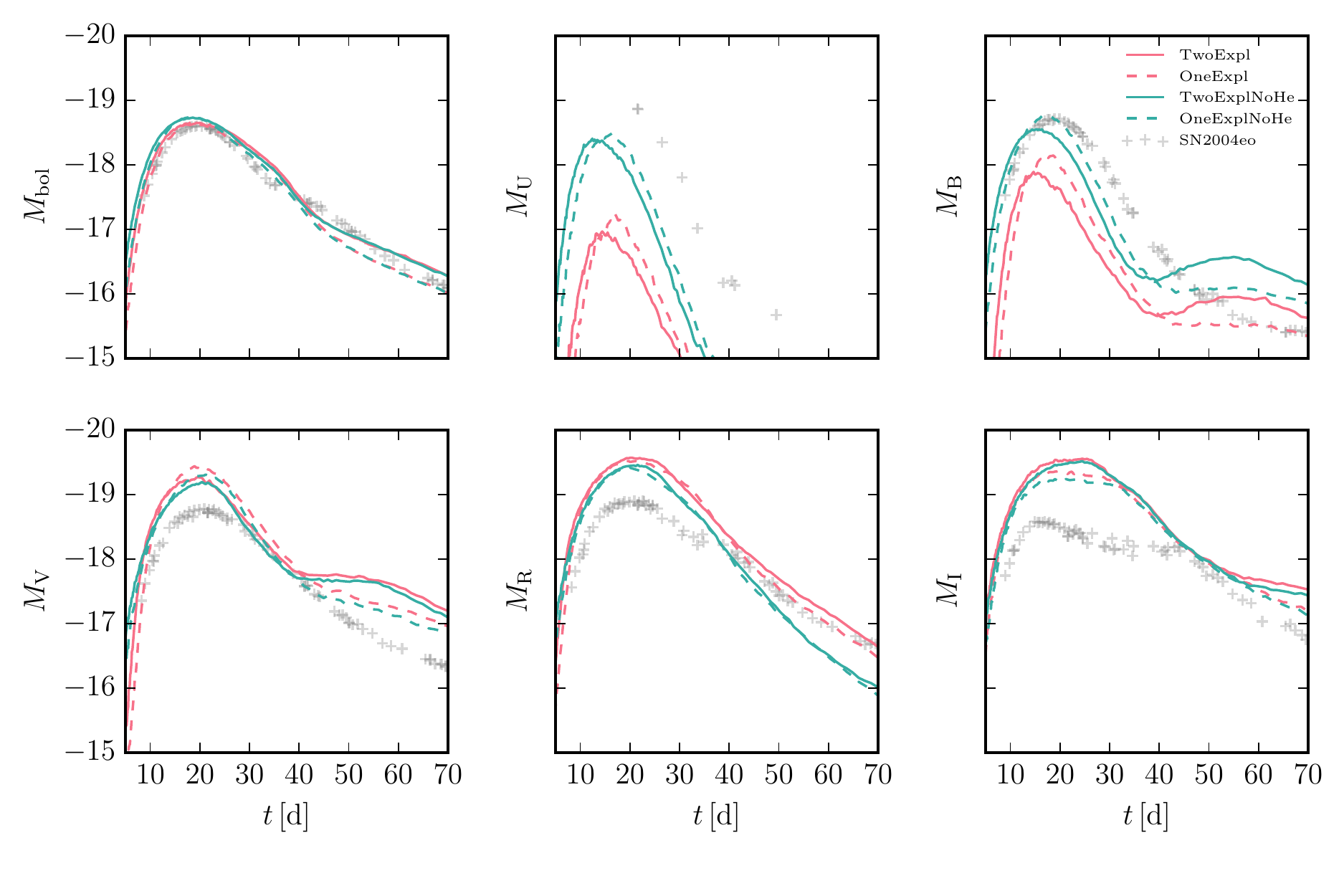}
\caption{Bolometric (top left panel) and  filtered light curves in the $U\!BV\!RI$ bands for the full model with surviving secondary WD (dashed red line) and the model with both white dwarfs exploding (solid red line). The cyan lines show the same models where we removed the ashes of the helium shells of both white dwarfs. The grey crosses show data of SN~2004eo \citep{Pastorello2007}. The bolometric light curve of SN~2004eo includes all data from $U$ to $K$ band. The light curves of the OneExpl and TwoExpl models are similar until about $40\,\mathrm{d}$ after the explosion.}
\label{fig:lightcurves}
\end{figure*}

\section{Synthetic light curves and spectra}

\label{sec:observables}

\begin{figure*}
\includegraphics[width=0.97\linewidth]{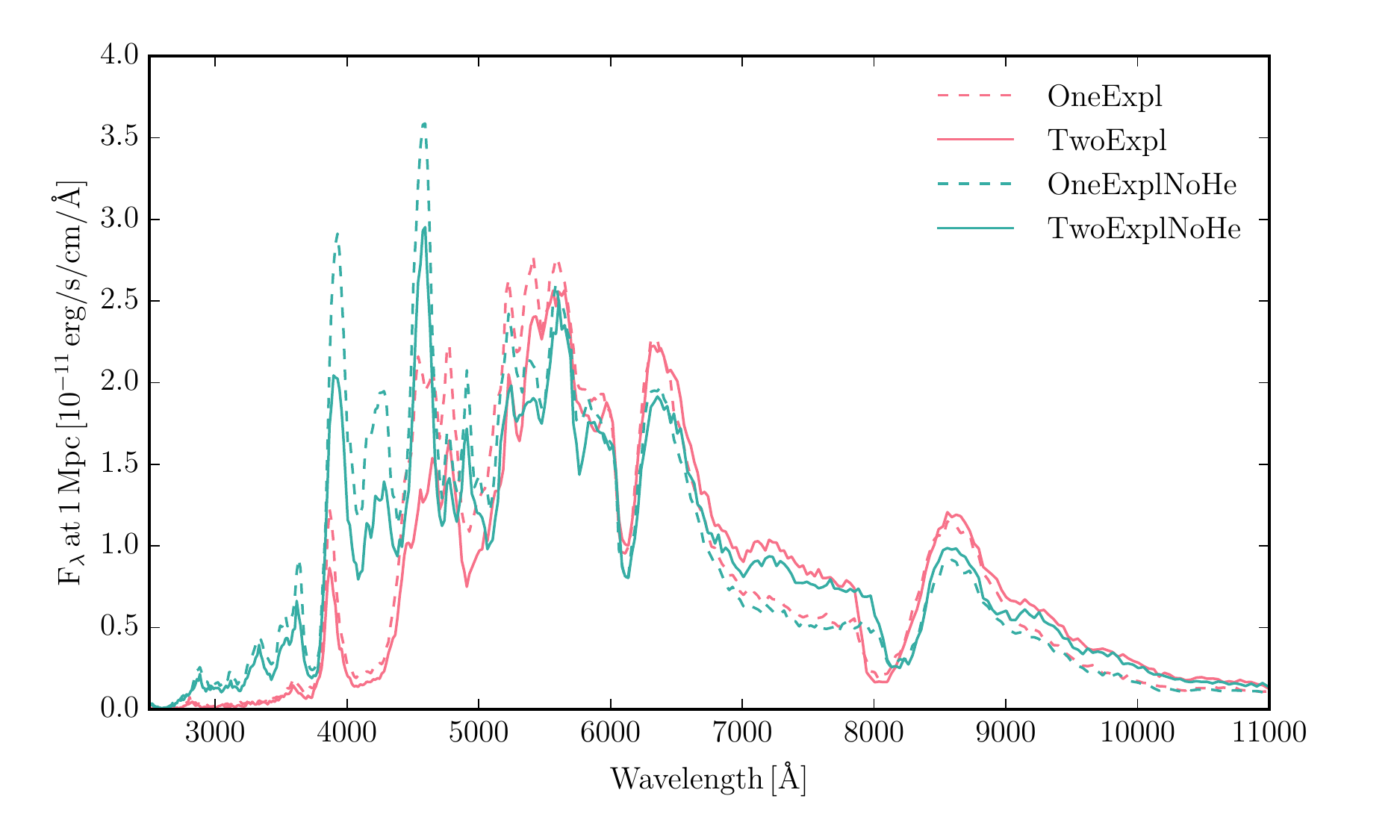}
\caption{Spectra at $t=20\,\mathrm{d}$ after the explosion. The red lines show the spectra for the full model with surviving secondary white dwarf (dashed line) and the model with both white dwarfs exploding (solid line). The cyan lines show the same models where we removed the ashes of the helium shells of both white dwarfs. The maximum-light spectra of the OneExpl and TwoExpl models are very similar but for a weak colour trend.}
\label{fig:spectra}
\end{figure*}

A proper comparison between models and observations requires forward modelling of the explosion models to synthetic observables and comparing those directly to observations. To this end, we show synthetic bolometric and filtered light curves for four different explosion models in Figure~\ref{fig:lightcurves}. Here we limit ourselves to 1D spherically symmetric radiation-transfer calculations to understand qualitative differences between our models and leave detailed 3D line-of-sight dependent synthetic observables to dedicated future studies.

We show the OneExpl and TwoExpl models described above. We also add two artificial models: the OneExpl and TwoExpl models from which the ashes of the helium shells have been removed from the ejecta by removing the tracer particles that have helium in their initial composition. We call these models OneExplNoHe and TwoExplNoHe. We also include observations of the well observed normal Type Ia supernova SN~2004eo \citep{Pastorello2007} which has a brightness similar to our models. Note that the small bumps in the synthetic light curves are Monte Carlo noise that we ignore in the discussion.

We focus on the differences between the two versions of OneExpl and TwoExpl runs, i.e., we are interested in predictions about how much of a difference the fate of the secondary white dwarf makes for observables. We find that the light curves of both pairs of models are very similar until $\sim\,40\,\mathrm{d}$ after the explosion. This is consistent with our expectation from the radial profiles discussed before. Only once the effective photosphere recedes below $v\lesssim 15\,000\,\mathrm{km\,s}^{-1}$, where the inner ejecta start to differ, can we see differences in the light curves. At later times, the TwoExpl and TwoExplNoHe models decline more slowly in the bolometric light curve than their counterparts, consistent with our estimate of $t_0$ in Sec.~\ref{sec:ejecta}.

The filtered light curves of the NoHe models are consistent with those of idealised models of centrally ignited isolated sub-Chandrasekhar-mass explosions \citep{Sim2010,Shen2018}. In our radiative-transfer calculations the models that include the helium shell are too red compared to observations owing to line blanketing by titanium and chromium produced in the burning of the helium shell of the primary white dwarf \citep{Kromer2010}. Importantly, however, the reddening of the light curves by the helium-shell ashes does not hide any differences between the light curves that arise from the core detonation in the secondary white dwarf. Moreover, a lower mass helium shell on the primary white dwarf and a more accurate full NLTE treatment of the helium shell may reduce the reddening. We conclude that the differences between OneExpl and TwoExpl models with and without the helium shell are small. The re-brightening of the TwoExpl models in the $\!B$ and $V$ bands after $40\,\mathrm{d}$ is likely a consequence of the approximate NLTE treatment of ionization in \textsc{artis} at such late times after explosion.
In the simulations, the Fe-group material recombines from doubly-ionised to singly-ionised, causing an increase in opacity leading to the re-brightening. However, we would expect that non-thermal effects would inhibit this recombination, and therefore the re-brightening may not be physical.
Nevertheless we show all light curves until $70\,\mathrm{d}$ after the explosion to emphasise the difference in the bolometric light curves that are unaffected by the NLTE assumptions.

We show spectra $20\,\mathrm{d}$ after the explosion close to maximum brightness for the same four models in Figure~\ref{fig:spectra}. We confirm that not only the filtered light curves but also the actual spectra for the OneExpl and TwoExpl models are very similar at this time. This statement again holds independently of the blanketing caused by the helium-shell ashes. So, importantly, the helium-shell ashes do not mask any otherwise observable differences. There is a slight tendency in both the filtered light curves and the spectra that the TwoExpl model is slightly redder than the OneExpl model, but this difference is small compared to typical differences between explosion models and between the models with and without helium-shell ashes.

\section{Discussion}

\label{sec:discussion}

Our explosion simulations have shown that the explosion of the secondary white dwarf can be completely hidden until long after maximum brightness, with no significant differences in observables until $\sim 40\,\mathrm{d}$ after the explosion.

In contrast to violent mergers that also completely destroy the binary system but lead to a very asymmetric explosion inconsistent with typical normal Type Ia supernovae \citep{Pakmor2012b,Bulla2016}, the explosion of the secondary white dwarf in the scenario discussed here does not lead to large asymmetries. Instead, the large-scale asymmetry of the ejecta is comparable to double-detonation explosions of single sub-Chandrasekhar-mass white dwarfs. This is mostly a result of the different timings of the carbon detonations: in the violent merger the merged object explodes all at once while in our new scenario the ejecta of the primary white dwarf expand for several seconds before the secondary white dwarf explodes.

Our scenario may provide an explanation for the apparent absence of a large number of surviving companions that would be seen as fast moving white dwarfs in the Milky Way \citep{Shen2018b}. These are otherwise predicted to exist if the double-degenerate channel with a surviving secondary white dwarf is the main channel for normal Type Ia supernovae. In principle, those surviving white dwarfs could be too faint to be observable \citep{Shen2018b}. However, they should still be visible directly in supernova remnants \citep{Kerzendorf2018,Shields2022}.

Importantly, the scenario in which the secondary explodes as well retains all desired population properties found for violent mergers. This includes the expected luminosity distribution \citep{Ruiter2013} and the ability to explain observed correlations between the luminosity of normal Type Ia supernovae and their host galaxies \citep{Kelly2010, Childress2013}.

In contrast to the model that only explodes the primary white dwarf, the model in which both white dwarfs explode has the potential to solve several discrepancies of late-time bolometric light curves. It shows a much more complicated 3D structure of the inner ejecta with notably almost no iron-group elements at velocities smaller than $5000\,\mathrm{km\,s^{-1}}$. We generally expect this as long as the mass of the secondary white dwarf is too low to produce iron-group elements when it explodes, i.e. roughly for secondary white dwarfs with $M<0.85\,\mathrm{M_\odot}$. In this way the TwoExpl model may explain observed nebular spectra of normal Type Ia supernovae, potentially even including the small number of objects that show bimodal structures in nebular emission lines \citep{Dong2015,Vallely2020}.

The asymmetry of the distribution of $^{56}\,\mathrm{Ni}$ and its large bulk velocity may help to explain large observed asymmetries \citep{Dong2018} and velocity offsets of emission lines in the nebular phase of Type Ia supernovae. In particular, velocity shifts larger than $2000\,\mathrm{km\,s}^{-1}$ \citep{Maeda2010} seem to be hard to explain with the orbital velocities of the binary system \citep{Shen2018b} or off-centre ignition in the white dwarf \citep{Chamulak2012} in sub-Chandrasekhar-mass explosions. Moreover, the explosion of the secondary in the TwoExpl model only changes the centre-of-mass velocity of the iron-group elements, but not that of the intermediate-mass elements of the ashes of the primary white dwarf. Thus it can increase the difference between the two. Finally, future 3D synthetic nebular spectra of the OneExpl and TwoExpl models and their differences will likely be able to either confirm or falsify our new model in which both white dwarfs explode.

Another route to distinguish the OneExpl and TwoExpl model may be to look at even later times and study nearby Type Ia supernova remnants. Again it seems necessary to compare them using full 3D models, that start to become feasible today \citep{Ferrand2022}. In particular it will be interesting to see if the different geometry of the central ejecta and different signature of the secondary white dwarf can be distinguished in synthetic observables.

The biggest uncertainty in our simulation very likely is the ignition of the carbon detonation in the secondary white dwarf. The strength of the detonation shock of the explosion of the primary white dwarf seems to help. Therefore we expect it to facilitate the detonation for more massive primary white dwarfs, because their explosions are more energetic and have a larger momentum. If the secondary white dwarf ignites only for a fraction of the exploding double-degenerate systems, we expect it will explode more often for more massive primary white dwarfs. However, other parameters like the mass of the secondary white dwarf and the mass of the helium shell on the secondary white dwarf can also influence the outcome and drastically increase the complexity of this scenario.

These parameters may determine whether or not the secondary white dwarf explodes at all. However, it is also possible that they only change the timing between the explosion of the primary white dwarf and that of the secondary white dwarf. For a smaller time lag between both explosions, we may expect the ejecta of the primary white dwarf to be more strongly affected, and the observables to show differences to the case with no explosion of the secondary earlier.

The second big uncertainty is the mass of the helium shell of the primary white dwarf at the time it starts explosive burning. This will depend on the formation history of the primary white dwarf as well as the amount of helium transferred from the secondary white dwarf to the primary white dwarf prior to the dynamical phase of the binary system that we can actually model. Even though the helium-shell masses used in this study seem to be possible in observed systems \citep{Pelisoli2021}, they are both highly uncertain and hard to model. So the obvious next step should be to do a parameter study that varies the helium-shell masses of both white dwarfs broadly for a given combination of white dwarf masses, i.e., we end up with four essentially independent parameters to cover.

Finally, we emphasize that it is crucial to simulate the full binary system and to take the secondary white dwarf fully into account to faithfully assess the validity of double-degenerate progenitor systems for Type Ia supernovae.

\section*{Data availability}
The simulations underlying this article will be shared on reasonable request to the corresponding author.

\section*{Acknowledgements}

CC acknowledges support by the European Research Council (ERC) under the European Union’s Horizon 2020 research and innovation program under grant agreement No. 759253. FPC acknowledges an STFC studentship and SAS acknowledges funding from STFC Grant Ref: ST/P000312/1. The work of FKR and AH  was supported by the Deutsche Forschungsgemeinschaft (DFG, German Research Foundation) -- Project-ID 138713538 -- SFB 881 (``The Milky Way System'', subproject A10), by the ChETEC COST Action (CA16117), and by the National Science Foundation under Grant No. OISE-1927130 (IReNA). FKR and AH acknowledge support by the Klaus Tschira Foundation. AH is a Fellow of the International Max Planck Research School for Astronomy and Cosmic Physics at the University of Heidelberg (IMPRS-HD). AJR acknowledges support from the Australian Research Council through award number FT170100243. ST has received funding from the European Research Council (ERC) under the European Union’s Horizon 2020 research and innovation program (LENSNOVA: grant agreement No 771776).




\bibliographystyle{mnras}


\bsp	
\label{lastpage}
\end{document}